\documentclass[useAMS,usenatbib,usegraphicx]{mn2e}
 
\newcommand\aj{AJ}%
\def\apj{ApJ}%
\def\apjl{ApJ}%
\def\aap{A\&A}%
\def\mnras{MNRAS}%
\def\ssr{Space~Sci.~Rev.}%
\title{A UV flux drop preceding the X-ray hard-to-soft state transition during the 2010 outburst of GX 339$-$4}

\author[Zhen Yan and Wenfei Yu]{Zhen Yan$^{1}$ and Wenfei Yu$^{1}$ \thanks{E-mail:wenfei@shao.ac.cn}\\
$^{1}$Key Laboratory for Research in Galaxies and Cosmology, Shanghai Astronomical Observatory,Chinese Academy of Sciences, \\
80 Nandan Road, Shanghai 200030, China } 

\begin{document}

\date{Accepted . Received ... ; in original form }

\pagerange{\pageref{firstpage}--\pageref{lastpage}} \pubyear{2011}

\maketitle
\label{firstpage}

\begin{abstract}
The black hole X-ray transient GX 339$-$4 was observed with  the {\it Swift} satellite across the hard-to-soft state transition during its 2010 outburst.  The ultraviolet (UV) flux measured with the filter UVW2 of the {\it Swift}/UVOT started to   decrease  nearly 10 days before  the drop in the hard X-ray flux when  the hard-to-soft state transition started. The UV flux $F_\mathrm{UV}$ correlated with the X-ray flux $F_\mathrm{X}$ as $F_\mathrm{UV}\propto F_\mathrm{X}^{0.50\pm0.04}$ before the drop in the UV flux. During the UV drop  lasting about 16 days, the X-ray flux in 0.4--10 keV was increasing.  The drop in the UV flux  indicates that  the jet started to quench 10 days before the hard-to-soft state transition seen in X-rays, which is unexpected. 
\end{abstract}
 
\begin{keywords}
accretion, accretion disks --- black hole physics ---ISM: jets and outflows--- X-rays:binaries --- X-rays:individual (GX 339$-$4)
\end{keywords}
 
\section{Introduction}
Most known black hole (BH) X-ray binaries (XRB) are transient sources. They usually stay in quiescence and occasionally undergo  dramatic X-ray outbursts.  These BH transients are known to experience different X-ray spectral states with particular X-ray spectral, timing  and radio properties during an outburst \citep{mr06},  including  two major spectral states. One is called the  soft state, the X-ray spectrum of which is dominated by a  multi-color blackbody  from an accretion disk \citep{ss73}.   The other is called the  hard state, the X-ray spectrum  of which is dominated by a powerlaw  with a photon index of $1.5<\Gamma<2.1$ \citep{mr06}.  Radio emission with a flat spectrum, which is thought as the self-absorbed synchrotron emission from a compact steady jet, is also a characteristic of the hard state \citep{fender01}. Suppressed radio emission in the soft state is thought to indicate the jet is quenched \citep[e.g.][]{fender99}. During the transition from the hard state to the soft state, the X-ray spectral properties are rather transitional and complex. This state is called  the intermediate state or the steep powerlaw state, the X-ray spectrum of which is composed of a steep powerlaw component ($\Gamma \geq 2.4$) sometimes combined with a comparable disk component \citep{mr06}. Many studies attempted to explain spectral state transitions   \citep[e.g.][]{esin97}, but to date there are still discrepancies between theories and observations probably due to the limitation of the underlying stationary accretion assumption, which is inconsistent with the observational evidence that non-stationary accretion plays a dominating role in determination of the luminosity threshold at which the hard-to-soft state transition occurs \citep{yy09}. 

In the hard state, the radio flux and X-ray flux, if not considering outliers, display a correlation in a form of $L_\mathrm{radio}\propto L_\mathrm{X}^{0.5-0.7}$\citep[e.g.][]  {corbel00,corbel03,corbel08,gallo03}, indicating  the coupling between the jet and the accretion flow.   A similar correlation also holds between the  optical/near-infrared (NIR) and   the X-ray flux \citep{homan05,russell06,coriat09},  which suggests that the   optical/NIR emission has the same origin as that of the radio emission,  i.e. the optically thick part of the jet spectrum can extend to NIR or even optical band \citep{cf02,bb04,homan05,russell06}.   In contrast to the extensive studies of the relations between   the radio, optical/NIR emission and the X-ray emission, there are much less studies about the ultraviolet (UV) emission in different X-ray spectral states because of limited UV observations.  {\it Swift}/UVOT has three filters to cover the UV band.  Simultaneous observations with the UVOT and the XRT onboard {\it Swift} provide unprecedented opportunities for us to study the UV emission in different X-ray spectral states.   

GX 339$-$4 is one of the best-studied BH X-ray transients which undergoes frequent outbursts in the past decades.  Because its donor star is very faint \citep{shahbaz01}, GX 339$-$4 during outbursts is therefore an excellent target   to investigate  the  UV/optical/NIR emission from the accretion and ejection in the various X-ray spectral states in detail.  In this paper, we focus on reporting the results obtained from our proposed  {\it Swift} observations of GX 339$-$4 during the 2010 outburst. We found the UV flux decreased well before the hard-to-soft state transition occurred.

\section{Observations and data reduction} 
In the past two decades GX 339$-$4 have undergone rather frequent outbursts. There is a unique correlation found between the peak flux in the hard X-ray  and the outburst waiting time in this source \citep{yu07,wy10}. This empirical relation allowed us to predict the peak  flux in the hard X-ray in an on-going outburst if the time when the hard X-ray flux reaches its peak can be determined \citep{wy10}. GX 339$-$4 entered a new outburst since the beginning of 2010 \citep{ya10,tomsick10,lewis10}, and then experienced an extended flux plateau of about 50 days which was unusual.  The accretion of disk mass during the extended plateau period was expected to cause the actual hard X-ray peak lower than the prediction. However  we were still able to predict the time of the hard-to-soft state transition based on a simple linear extrapolation of the rising trend and the relation predicted in \citet{wy10}. We requested a series of target of opportunity (TOO) observations with {\it Swift} spanned by about 20 days, which indeed covered the  hard-to-soft  state transition.    These observations, together with other TOO observations immediately before or after (under Targets IDs 30943,31687) were analysed, which covered the period from January 21, 2010 to June 19, 2010 (MJD 55217--55366).  

Due to high count rates, all the XRT observations of GX 339$-$4 during this period were automatically taken in the windowed timing (WT)  mode which only provided one-dimensional imaging. We processed all the initial event data with the task {\it xrtpipeline} to generate cleaned event files by standard quality cuts for the observations. There was no cleaned event data generated for  the observation  00030943009, so we excluded this observation in our XRT data analysis. We extracted all the cleaned WT mode events using XSELECT v2.4 as part of the HEASOFT v6.11 package.  Source extraction was performed with a rectangular region of 20 pixels $\times$ 40 pixels centered  at the coordinate of GX 339$-$4 and the background extraction was performed with two rectangular regions of 20 pixels $\times$20 pixels whose centers are 50 pixels away from the coordinate of GX 339$-$4. Some observations with high count rates  ($>$ 150 c/s) are affected by the pile-up.  The easiest way to avoid  distortions produced by the pile-up is to extract spectra by eliminating the events in the core of the point spread function (PSF). The method by investigating the grade distribution \citep{pagani06,mineo06} was used to determine the excluded region,  which was chosen as the central 2 pixels$\times$20 pixels, 4 pixels$\times$20 pixels, 6 pixels$\times$20 pixels and 8 pixels$\times$20 pixels when the count rate was in the range of 150--200 c/s, 200--300 c/s, 300--400 c/s and 400--500 c/s, respectively.   After events selection,  we produced the ancillary response files (ARFs) by the task {\it xrtmkarf}. The process included inputting exposure maps with {\it xrtexpomap} and applying the vignetting and PSF corrections. The standard redistribution matrix file (RMF, {\it swxwt0to2s6\_20010101v014.rmf}) in the CALDB database was used for the spectral analysis. All the spectra were grouped to at least 20 counts bin$^{-1}$ using {\it grppha} to ensure valid results using $\chi^{2}$ statistical analysis.

 {\it Swift}/UVOT  has six filters: V, B, U, UVW1, UVM2 and UVW2.    We applied  aspect corrections to all the UVOT data by using {\it uvotskycorr}.  For each observation, the sky image of each filter was summed using {\it uvotimsum} in order to increase photon statistics, then the spectral files which are compatible with XSPEC can be created by {\it uvot2pha}. We performed aperture photometry with the summed images by using  {\it uvotsource} with an aperture radius 5$^{\arcsec}$. The background region was chosen in a larger void region, which is away from the target. The photometry results in AB magnitude system are shown in the upper panel of Fig.~\ref{f2}.

\section{Data analysis and results}
\begin{figure}
\centering
\includegraphics[width=0.5\textwidth]{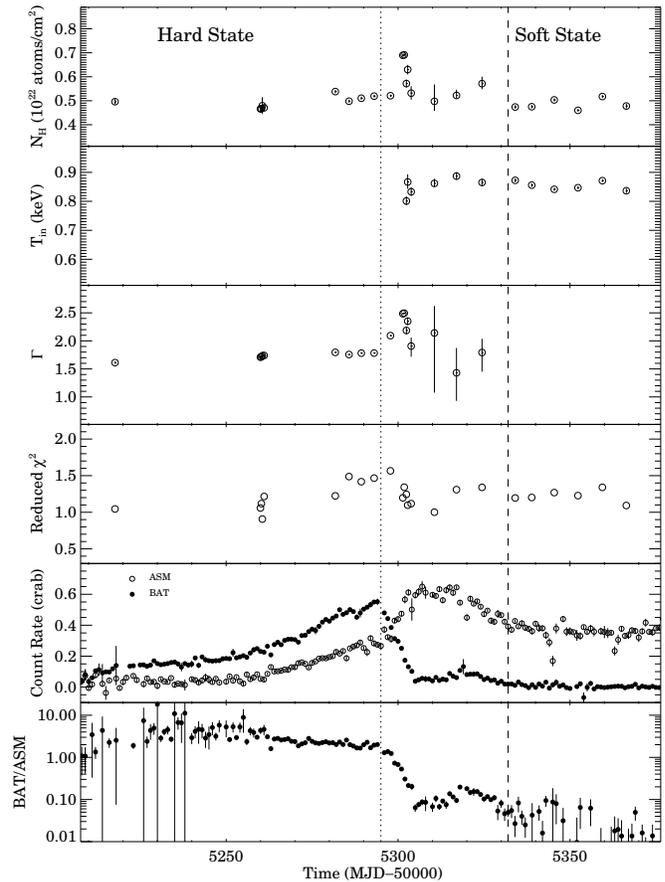}
\caption{From top to bottom, {\it Swift}/XRT spectral parameters column density $N_{H}$, inner disk temperature $T_{in}$, photon index,  reduced $\chi^{2}$, the {\it RXTE}/ASM and {\it Swift}/BAT lightcurves, and the hardness ratio between the  BAT and the ASM fluxes.  The vertical dotted  and dashed  lines indicate the time interval when the source transited between the hard and the soft state.}
\label{f1}
\end{figure} 

\begin{figure}
\centering
\includegraphics[width=0.5\textwidth]{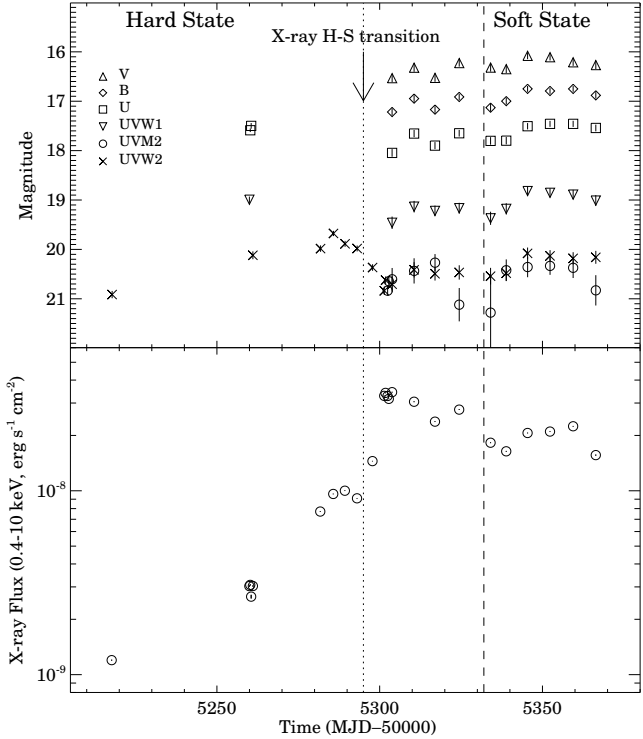}
\caption{The X-ray and UV/optical lightcurves of GX339$-$4 across the hard-to-soft state transition during the 2010 outburst.  Spectral state boundaries are indicated by the vertical lines.  The UVOT photometry results are plotted in the upper panel. The UVW2 filter was used through all the observations. The UV flux initially increased with the X-ray flux, and  then decreased   after MJD 55285,  while the X-ray flux was still increasing.}
\label{f2}
\end{figure}

All the XRT spectra were fitted in XSPEC v12.7.0.  The photons with energies below 0.4 keV were ignored during the spectral fits because of hot pixel problems at lower energy (Kim Page, private communication). The  0.4--10 keV X-ray spectra during the period of MJD 55217--55293 can be fitted well by a model composed of an absorbed powerlaw with an photon index $\sim$ 1.61--1.79, which indicates the source was in the hard state. Then the photon index steepened ($\Gamma \sim$ 2.09--2.49)  in the period of MJD 55297--55324 and an additional disk component was required after MJD 55302, indicating the source was leaving from the hard state to the soft state \citep[see also ][]{motta10a}.   The fitted $N_{H}$ shows a sudden increase during the state transition. A possible reason for the high absorption is that the powerlaw model does not exhibit a low energy cut-off  which is caused by the seed photon of the Comptonization. We fitted the XRT spectra in order to get the photon index and to estimate the X-ray flux, and the discussion of the physical spectral components is beyond the scope of this paper. So we used the simple powerlaw model rather than the Comptonization models for simplicity. The 0.4--10 keV X-ray spectra taken after MJD 55334 can be fitted with only an absorbed multi-color disk blackbody model ($T_{in} \sim$ 0.84--0.87 keV), indicating that the source has  entered the soft state.  The best-fit spectral parameters are plotted in Fig.~\ref{f1}.  These model parameters were used to show the spectral states the source resided in (see Fig.~\ref{f1}) and to estimate the 0.4--10 keV X-ray flux (see Fig.~\ref{f2}). We also plotted the public daily lightcurves of GX 339$-$4 obtained from  the {\it RXTE}/ASM and the {\it Swift}/BAT in Fig.~\ref{f1}. The X-ray spectral states are distinguishable by the hardness ratio between the count rates in the BAT (15--50 keV)  and in the ASM (2--12 keV)  \citep[see also ][]{yy09,tang11}, consistent with the XRT spectral results (see Fig.~\ref{f1}).

 The upper panel of Fig.~\ref{f2} shows the magnitudes of six UVOT filters in different X-ray spectral states. The UVW2 filter was used throughout all the observations. The flux in the UVW2 band increased with the X-ray flux at the beginning, and then began to decline about 10 days before the hard-to-soft state transition started. The UVW2  flux decreased  by   about 62\% in a period of 16 days since MJD 55285 and then kept at a steady level (see Fig.~\ref{f2}).   Yale SMARTS XRB project \footnote {http://www.astro.yale.edu/buxton/GX339/} has offered optical/NIR photometric coverage of this outburst on a daily basis. The optical/NIR flux also showed a drop \citep[see also][]{russell10,cb11,buxton12}, which was  simultaneous with the UV flux drop within the uncertainties of the sampling.  The flux decreased by about 80\% in V band, 84\% in I band, 92\% in J band and 94\% in H band, respectively.  As we can see, the decreased amplitudes were wavelength dependent, i.e. the flux decreased more at lower frequencies than at higher ones \citep[see also ][]{homan05}.  Moreover, such a drop in radio flux was also seen \citep{zhang12}, implying a common physical origin  for the flux drops in the broad bands. 

\begin{figure}
\centering
\includegraphics[width=0.5\textwidth]{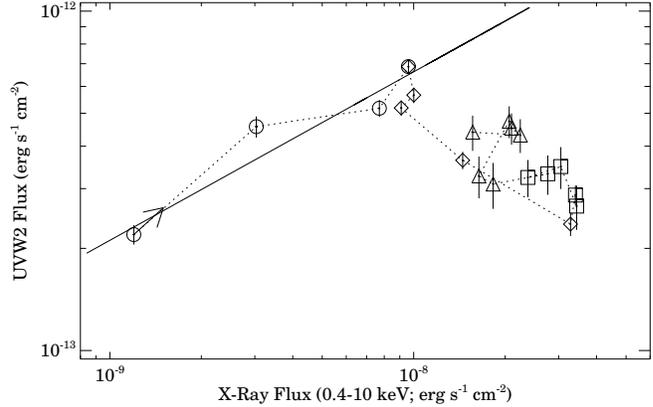}
\caption{UVW2 band flux vs. 0.4-10 keV X-ray flux. The circles represent the data when the UV flux increased,  the diamonds represent the data when the UV flux decreased,  the triangle represent the data in the soft state, and the squares represent the rest of data. The data with maximal UV flux is the end of UV flux increase and the start of the UV flux decrease, so two symbols (a circle and a diamond) were plotted on this data.  The solid line shows the best power-law fit to the data during UV flux increasing (circle symbols) with an index of $0.50 \pm 0.04$. The dotted line shows how the X-ray and UV fluxes evolve, and the arrow indicates the beginning.}
\label{f3}
\end{figure}

We further investigated how the UV flux correlated with the X-ray flux in the hard state and across the hard-to-soft state transition during the 2010 outburst of GX 339$-$4 (see Fig.~\ref{f3}). The UV flux correlated with the unabsorbed 0.4--10 keV X-ray flux in a power-law form with an index of $\sim 0.50\pm0.04$ before its drop. The X-ray flux and the UV flux in the soft state did not follow this correlation (see Fig.~\ref{f3}), indicating that the origins of the UV emission differ between the hard state and the soft state.

\section{Discussion}
The origin of the UV emission of black hole X-ray binaries in the hard state has not been well-understood.   The UV emission could be from the jet, the cool disk, the hot ADAF and X-ray reprocessing in the accretion disk  \citep{markoff03,yuan05,rykoff07,maitra09}. Let's look at these options in details. The correlation of the form of $F_\mathrm{UV}\propto F_\mathrm{X}^{0.50 \pm 0.04}$ (see Fig.~\ref{f3}) before the UV flux drop is quantitatively inconsistent with a disk origin of the UV emission, since the UV flux $F_\mathrm{UV}$ should be proportional to the X-ray flux as $F_\mathrm{X}^{0.3}$  for a simple viscously heated disk \citep[see ][]{russell06}. X-ray reprocessing in the accretion disk could be a promising option if not considering the UV flux drop before the hard-to-soft transition we report in the paper.  \citet{paradijs94} has predicted the emission of X-ray reprocessing should be proportional to $L_\mathrm{X}^{0.5}$, which was seen in  XTE J1817$-$330 during the decay of the 2006 outburst in which the UV luminosity was found to correlate with the 2--10 keV X-ray luminosity as $L_{X}^\alpha$ where $\alpha=0.47\pm0.02$ \citep{rykoff07}. In the case of GX 339$-$4 the UV flux drop was accompanied by an X-ray increase, against the X-ray reprocessing interpretation.  Therefore the origin of the UV emission before the UV flux drop as X-ray reprocessing by the outer disk is also ruled out. 

The flux drop in the radio or optical/NIR band usually associated with the hard-to-soft state transition is  thought due to the jet quench  \citep{fender99,homan05,coriat09}.  In the observations we analyzed, a drop of the UV flux started about 10 days before the transition. The flux drop is similar to those drops in the radio, the NIR, and the optical, which is highly suggestive of a common origin. So the drop in the UV flux should be due to jet quenching as well.  
 
After the jet quenched, the UV/optical/NIR fluxes declined to a steady flux level, indicating additional  components other than the compact jet contributed to the UV/optical/NIR emission. The likely source is the X-ray irradiation, since the steady UV/optical/NIR fluxes maintain through the soft state, and we found the irradiation model {\it diskir} \citep{gier09} can provide an acceptable fit to the UVOT and XRT data in the soft state.  Moreover when the hard X-ray reached its maximum, the UV/optical emission was at a very low level (see Fig.~\ref{f2}), indicating potential X-ray irradiation contributes little UV/optical/NIR emission.  Therefore,  the dominant UV/optical/NIR emission during the rising phase of the outburst was from the compact jet.  

The scales of the flux densities corresponding to the drops in the  UV/optical/NIR bands  (peak flux density minus the steady flux density after the jet quenching)  should be all contributed by the compact jet, if the X-ray irradiation also contributed a steady level of UV/optical/NIR flux densities during the jet quenching. Then the scales of the dereddened flux densities corresponding to the drops in the UV/optical/NIR bands can be used to construct a spectral energy distribution (SED) of the emission from the compact jet. However, the observed SED in UV/optical/NIR bands is highly influenced by the extinction, which is not well constrained in GX 339$-$4. We then used different color excesses  to deredden the  scales of the flux densities corresponding to the drops in the UV/optical/NIR bands with the  extinction law of \citet{cardelli89}. To convert the magnitudes obtained from SMARTS to flux densities, we used the zero-point fluxes in \citet{bessell98}. We then constructed the SED in the wavelength from radio to UV/optical/NIR in order to investigate the jet spectrum.  The  flux density of UVW2 band reached its peak on MJD 55285, the closest radio observation (on MJD 55286) shows the radio flux densities were 22.38 mJy and 25.07 mJy at 5.5 GHz and 9 GHz, respectively \citep{zhang12}.  The two radio flux densities can be well fitted by a power-law with  an index $\sim$ 0.19, which is consistent with the optically thick synchrotron emission of the jet.  If using the color excess of $E(B-V) = 1.2 $ given by \citet{zdz98},  the extrapolation of the radio spectrum coincides with the  scale of the dereddened flux density corresponding to the drop in the UVW2 band, indicating that  the optically thick spectrum of the jet might had extended to the UVW2 band.   But the scales of the dereddened flux densities corresponding to the drops in the V, I, J and H bands are all below the extrapolation of the radio spectrum. We found the UV-NIR SED can be well fitted with two power-law components. From the H band to the  I band, the power-law index was about $-0.27$, and from the I band to the UVW2 band, the power-law index is about $+1.16$.   The data in different wavelengths were not exactly simultaneous.  So this complex SED may be caused by the large variation of the jet spectrum on the timescale of less than one day in the optical/NIR band \citep{gandhi11}.  If using the color excess of  $E(B-V) = 0.94$ given by \citet{homan05},  the scales of  the dereddened flux densities corresponding to the drops in the UV/optical/NIR bands were all below the extrapolation of the radio spectrum and also show a turnover feature in the I band. The turnover feature can be well fitted  with two power-law components with the power-law index  $\sim -0.66$ from H band to I band, and $\sim +0.16$ from the I band to the UVW2 band. If using the much smaller color excess of $E(B-V) =0.66$ given by \citet{motch85}, the scales of  the dereddened flux densities corresponding to the drop in the UV/optical/NIR bands can be well fitted  by a single power-law with an index of $-1.06$, which is roughly consistent with the optically thin part of the jet spectrum.  Therefore, different color excess leads to different conclusion about the jet spectral properties.  The jet spectrum can not be well constrained in the UV/optical/NIR bands due to large uncertainties in the estimation of the extinction. 

However, because there is a positive correlation between the UV  and  the X-ray fluxes during the UV flux rise ($F_\mathrm{UV}\propto F_\mathrm{X}^{0.50 \pm 0.04}$, see also Fig.~\ref{f3}),  we suggest that  the optically thick part of the jet spectrum could extend to the UVW2 band \citep[see, e.g., the discussion in ][]{coriat09}.  If this is true, the break frequency between the optically thick and the optically thin parts of the jet spectrum should have been in the UVW2 band or higher frequencies.  Then a lower limit of $-1.17$ on the spectral slope of the optically thin synchrotron emission of the jet can be given by a power-law fit from the UVW2 band to 0.4 keV, which is steeper than the typical spectral slope ($\sim -0.7$) of the optically thin synchrotron jet emission. Spectral slopes steeper than $-0.7$ have been found before, for example,  $-0.95$ in GX 339$-$4 \citep{dincer12} and $-1.5$ in XTE J1550$-$564 \citep{russell2010}. 

If the UV flux is mainly from optically thick synchrotron emission of the compact jet, the jet should be very energetic. We estimated the radiative flux of the jet at the UV peak was  $3.21\times 10^{-9}~\mathrm{ergs}~\mathrm{s}^{-1}~\mathrm{cm}^{2}$ by extrapolating the radio spectrum from 5.5 GHz to UVW2 band.  The corresponding 0.1-100 keV X-ray flux, extrapolated from the 0.4--10 keV spectrum, was about $2.53 \times 10^{-8}~\mathrm{ergs}~\mathrm{s}^{-1}~\mathrm{cm}^{2}$, corresponding to $\sim 0.15~L_\mathrm{Edd}$ if adopting a distance of 8 kpc and a mass of $10~M_{\sun}$ \citep{zdz04}. Taking reasonable radiative efficiency of the jet of about 0.05 and the Doppler correction factor of about 1 \citep{fender01,cf02}, the jet power $P_\mathrm{jet}$ is about 2.5 times of the X-ray luminosity $L_\mathrm{X}$. This is larger than previous estimates of $P_\mathrm{jet}/L_\mathrm{X}$ in the hard state of GX 339$-$4, which were all less than 1  \citep[e.g. ][]{cf02,dincer12}.    Moreover, the jet power at the UV flux peak was greater than the X-ray luminosity during a bright hard state of more than 0.1  $L_\mathrm{Edd}$, much higher than the critical value of $4\times 10^{-5} L_\mathrm{Edd}$ determined in \citet{fender03}. It is worth noting that a much lower jet power ( $P_\mathrm{jet}/L_\mathrm{X} \sim 0.05$) was seen in GX 339$-$4 at similar X-ray luminosity $\sim 0.12~L_\mathrm{Edd}$ \citep{cf02}. Therefore similar X-ray luminosity corresponds to a large range of jet power in the same source. This phenomenon is similar to the hysteresis effect of spectral state transitions, in which mass accretion rate is not the dominant parameter \citep{yy09}. This suggests that the jet power might be determined by non-stationary accretion, which leads to insignificant dependence on the mass accretion rate or X-ray luminosity. 

\section*{acknowledgements}
We would like to thank the anonymous referees for valuable comments and suggestions.  We thank Kim Page, Chris Done, Albert Kong and Wenda Zhang for help on the {\it Swift}/XRT data analysis, Andrzej A. Zdziarski and Feng Yuan for useful discussions and the {\it Swift} team for the scheduling the TOO observations.  We also acknowledge the use of data obtained through the High Energy Astrophysics Science Archive Research Center Online Service, provided by the NASA/Goddard Space Flight Center. This work is supported by the National Natural Science Foundation of China (10773023, 10833002,11073043), the One Hundred Talents project of the Chinese Academy of Sciences,  the National Basic Research Program of China (973 project 2009CB824800), and the starting funds at the Shanghai Astronomical Observatory.


 \label{lastpage}
\end{document}